\documentclass[twocolumn,           
               showpacs,            
               nopreprintnumbers,     
               aps,                 
               prd,          	    
               letterpaper,             
               groupeaddress,      
               nofootinbib,         
               tightenlines,        
               floats,floatfix      
               ]{revtex4-1}

\usepackage{graphicx}
\usepackage{fancyhdr}

\usepackage[english,spanish]{babel}

\usepackage[applemac]{inputenc}

\usepackage[T1]{fontenc}

\catcode`¿
\catcode`á
\catcode`é=13
\catcode`í=13
\catcode`ó=13
\catcode`ú=13
\catcode`ü=13
\catcode`ñ=13
\catcode`Á=13
\catcode`É=13
\catcode`Í=13
\catcode`Ó=13
\catcode`Ú=13

\def¿{?`}
\defá{\'a}
\defé{\'e}
\defí{\'i}
\defó{\'o}
\defú{\'u}
\defü{\"u}
\defñ{\~n}
\defÁ{\'A}
\defÉ{\'E}
\defÍ{\'I}
\defÓ{\'O}
\defÚ{\'U}

\def·{\'a}
\defÈ{\'e}
\defÌ{\'i}
\defÛ{\'o}
\def˙{\'u}
\defÒ{\~n}
\def¡{\'A}
\def…{\ldots}
\def°{!`}
\def¡{!`}

\usepackage{amsmath}
\usepackage{amsfonts}
\usepackage{amssymb}
\usepackage{graphicx}
\usepackage{gensymb}
\usepackage{bm}		
\usepackage{subfigure}
\usepackage{amsfonts}
\usepackage{booktabs}
\usepackage{float} 

\graphicspath{ {./}{figures/}}

\begin{document}

\title{Los agujeros negros y las ondas del Doctor Einstein}

\author{Mario A. Rodr\'iguez-Meza}
\email{marioalberto.rodriguez@inin.gob.mx (Corresponding author)}
  
\address{Departamento de F\'{\i}sica, Instituto Nacional de
Investigaciones Nucleares, 
Apdo. Postal 18-1027, M\'{e}xico D.F. 11801,
M\'{e}xico}

\begin{abstract}
We describe the main scientific developments that lead LIGO project to the
detection of the gravitational waves: general relativity, black holes and gravitational
waves predictions; numerical relativity and the collision and coalescence simulations
of binary black holes and the development of different kind of gravitational wave
detectors. Most important, this detection is confirming the existence of the enigmatic
black holes.


\bigskip
Se describen los principales desarrollos científicos que llevaron a la
detección de las ondas gravitacionales por el proyecto LIGO: la relatividad general, sus
predicciones de agujeros negros y ondas gravitacionales, la relatividad numérica y las
simulaciones de la colisión y coalescencia de dos agujeros negros y los desarrollos de
diversos detectores de ondas gravitacionales. Finalmente la detección de las ondas
gravitacionales confirma la existencia de los enigmáticos agujeros negros.

\medskip
\noindent
Palabras clave: relatividad general, ondas gravitacionales, agujeros negros, relatividad
numérica, interferómetro.
\end{abstract}

\date{\today}
\maketitle

\section{Introduction}\label{sec:Intro}
Podríamos comenzar este relato contando lo común y
corriente cuando se habla de la relatividad general, o que
Albert Einstein presentó su versión hace poco más de cien
años (25 de noviembre de 1915) ante la Academia Prusiana
de Ciencias en Berlín. Sin embargo, lo que ya no es muy
conocido es que cinco días antes de que Einstein presentara
su “versión”, el matemático David Hilbert le había \emph{ganado}
al presentar otra versión, en Göttingen, la cual dedujo a
partir de un principio variacional (Rodríguez-Meza, 2015).

Meses después de estos dos reportes, en 1916, el astrónomo
alemán Karl Schwarzschild publica la primera solución a las
ecuaciones, bastante complicadas, de la relatividad general. La
solución que encontró Schwarzschild es la precursora de los
enigmáticos agujeros negros, esos objetos que atrapan todo
lo que está a su alcance, incluso la luz, de ahí su nombre. Y
también en 1916 Einstein encontró soluciones tipo ondulatorio
a su ecuaciones linealizadas, es decir, en el régimen de
campo débil donde se podrían encontrar algunos resultados
(Einstein, 1916). Encontró que el espacio y el tiempo pueden
oscilar propagando ondas gravitacionales que se mueven a la
velocidad de la luz. Estas ondas son perturbaciones pequeñísimas
sobre las que hubo poca atención hasta el debate en
1957, en Chapel Hill, North Carolina, en el que se puso sobre
la mesa la pregunta sobre su existencia y posibilidad de ser
medidas. Es justo la perturbación gravitacional ondulatoria
producida por la colisión dramática de dos agujeros negros
masivos lo que las antenas gemelas de LIGO (Laser Interferometer
Gravitational-wave Observatory) detectaron el 14
de septiembre de 2015 (Abbott \emph{et al}., 2016). LIGO no estaba
oficialmente en funciones, puesto que había sido apagado
en 2010. Después, cuando los interferómetros habían sido
mejorados en cuanto a la sensibilidad de los detectores,
aumentando la potencia de los láseres y mejorando el aislamiento
sísmico de los espejos, los técnicos hicieron pruebas
y arrancarían en unos días más; no obstante, de pronto
apareció la señal y \emph{escucharon el gorjeo}. El anuncio de este
hallazgo espectacular se acordó que se haría para el mismo
día en que la revista \emph{Physical Review Letter} publicara el artículo
con los resultados, lo cual sucedió el 11 de febrero de 2016
(Abbott \emph{et al}., 2016). Bien, comencemos.


\section{Relatividad general}\label{sec:RG}

¿Qué es la relatividad general? Esta teoría nos dice que la
gravedad ya no es más una fuerza como Newton lo pensaba,
ahora es una deformación del espacio y el tiempo (que
fueron unificados por el matemático Hermann Minkowski
en una sola entidad, el espacio-tiempo) producida por los
cuerpos. Por ejemplo, el Sol deforma el espacio-tiempo
en su vecindad y los planetas se mueven siguiendo las
geodésicas de este espacio-tiempo curvado, trayectorias
con la longitud mínima de entre las posibles que unen dos
puntos en este espacio-tiempo. La idea de Einstein fue
darse cuenta ---al igual que en la relatividad especial que
había formulado en 1905--- que con sólo dos postulados
podría desarrollar la teoría: a) las leyes son las mismas sin
importar el observador (esté en movimiento uniforme o
en movimiento acelerado) y b) la gravedad y la aceleración
son equivalentes. El primero es el principio de relatividad
del movimiento no sólo aplicable a observadores inerciales
sino también a los acelerados, también conocido como
principio de covariancia general, mientras que el segundo
postulado es el principio de equivalencia y Einstein llegó a
él usando sus famosos experimentos pensados. Julio Verne
en su novela \emph{Alrededor de la Luna}, publicada en 1869 ---100
años antes del alunizaje del Apolo 11---, ya había anticipado
esta idea de equivalencia en la escena de \emph{Satélite}, el perrito
muerto. Julio Verne muere en 1905, el \emph{annus mirabilis} de
Einstein. Desarrollando más las ideas, Einstein concluyó
entonces que la gravedad es el resultado de que el espacio-tiempo
se deforma curvándose en presencia de los cuerpos.
Entre más masivo sea un objeto mayor es la deformación
del espacio-tiempo en la vecindad de él. Además, estableció
que la dinámica de los cuerpos es tal que siguen trayectorias
geodésicas de este espacio-tiempo curvo, lo que nos
lleva al problema de calcular distancias en espacios curvos
generales. En un espacio plano como aquel al que estamos
acostumbrados, este cálculo se hace empleando el teorema
de Pitágoras, que para dos puntos muy cercanos (infinitesimalmente
cercanos, dirían los matemáticos) la distancia se
escribe en coordenadas cartesianas como: $ds^2 = dx^2 + dy^2$.
Esta pequeña distancia, $ds$, se llama \emph{elemento de línea}. En
un sistema de coordenadas polares cambia a: 
$ds^2 = dr^2 + r^2d\theta^2$, 
el elemento de línea depende de la posición en este
caso. Si hubiera tan sólo un sistema de coordenadas en el
cual los términos cuadráticos diferenciales llevaran todos
coeficientes constantes, entonces el espacio será plano, sin
curvatura, que es este caso. En general, el cálculo de una
longitud en una superficie curva arbitraria se hace usando:
$ds^2 = g_{xx} dx^2 + g_{xy} dx dy + g_{yy} dy^2$. 
A los coeficientes $g_{xx}$,
$g_{xy}$, y $g_{yy}$, se les conoce como \emph{coeficientes métricos} y la relatividad
general nos dice cómo obtenerlos para el espacio-tiempo
en presencia de cuerpos masivos. En particular, en
la relatividad especial, Minkowski escribió el elemento de
línea como: 
$ds^2 = c^2 dt^2 - dx^2 - dy^2 - dz^2$. Para la luz, 
$ds =0$, 
que es consistente con que la velocidad de la luz es una
constante e igual a $c$.

Con esta metodología que desarrollaron Einstein (y
Hilbert) se pueden calcular las órbitas de los planetas alrededor
del Sol. Estas órbitas son prácticamente iguales a las
newtonianas en todos los casos, excepto para Mercurio, en
cuyo caso por su cercanía al Sol se manifiestan claramente
los efectos relativistas, fenómeno conocido como la \emph{precesión
del perihelio de Mercurio}. Las predicciones de la relatividad
general fueron a) corrimiento al rojo de la frecuencia de
una onda luminosa en presencia de un campo gravitatorio
intenso, b) deflección de la luz al pasar por la vecindad de
un campo gravitacional intenso como el del Sol y c) las
ondas gravitacionales que son pequeñas fluctuaciones del
espacio-tiempo. Confirmadas todas, excepto las ondas
gravitacionales, aunque se podría decir que ya fueron
detectadas indirectamente con el estudio observacional del
pulsar binario PSR1913+16, cuyo periodo va disminuyendo
al emitir ondas gravitacionales (Hulse y Taylor, 1975). Los
datos de la evolución temporal del periodo se ajustan casi
de manera perfecta con lo que predice la relatividad general.
Estas pruebas a la relatividad general son en el límite de
campo débil. No hay pruebas en el límite de campo fuerte,
hasta ahora.


\section{Agujeros negros}\label{sec:BH}
La solución encontrada por Schwarzschild fue otra de las
predicciones. Presentaba una singularidad, un defecto que
no era muy agradable, para la región interna limitada por
un radio, llamado de \emph{Schwarzschild}. No obstante, no se le
prestó mucha atención, ya que lo que importaba era la solución
más allá del radio de Schwarzschild donde no había
problemas. Pero se comenzó a entender mejor la evolución
y muerte de las estrellas. Una estrella al hacerse vieja se
va contrayendo y bajo ciertas circunstancias podría colapsarse
convirtiéndose en un objeto muy pequeño y denso.
Entonces, se pensó que este asunto molesto del radio de
Schwarzschild debía estudiarse en detalle y Einstein \emph{le
metió el diente} y trató de probar que una estrella no podría
colapsarse y alcanzar un radio menor al de Schwarzschild.
El modelo que construyó era muy simple, muy a su estilo
como su modelo de un sólido, todos los átomos vibrando
con la misma frecuencia y por lo simple no logró explicar
la capacidad calorífica de los sólidos a bajas temperaturas
(Cheng, 2013). En este caso, Einstein consideró que todas
las partículas seguían órbitas circulares en planos diferentes
pero todos con un centro común, el centro de masa del
sistema. La conclusión que obtuvo fue que las órbitas no
podrían ser menores que 1.5 veces el radio de Schwarzschild
y anunció que la singularidad de Schwarzschild no
tenía realidad física. Entonces, Oppenheimer y su estudiante
Hartland Snyder revisaron el problema y en vez de
las órbitas circulares de Einstein consideraron un sistema
más realista; un aglomerado esférico de neutrones a cuyo
colapso se le opone la fuerza nuclear repulsiva entre ellos.
Encontraron que si la masa de la configuración es mayor
que tres cuartos la masa del Sol, la fuerza nuclear no le gana
a la fuerza gravitacional atractiva y el sistema se colapsa y
además no se detiene en el radio de Schwarzschild sino que
se hunde por siempre. Este extraño comportamiento sólo
se logró entender hasta 1960 cuando los físicos se dieron
cuenta que la singularidad de Schwarzschild era un \emph{artifact}
matemático que resultaba de la mala elección del sistema
de coordenadas. La masa colapsada en realidad formaba
una región muy curvada del espacio-tiempo rodeada por
una superficie al radio de Schwarzschild dentro de la cual
nada escapa. John Wheeler, en 1967, le dio el nombre de
\emph{agujero negro} a estos extraños objetos. El nombre era trivial,
pero resultó todo un acierto “comercial”. Es un objeto
masivo colapsado, como una estrella que bajo ciertas
condiciones se colapsa, sin dejar rastro ni de su materia,
ni de sus manchas y prominencias, ha desaparecido como
el gato Cheshire en \emph{Alicia en el país de las maravillas} 
dejando
tan sólo la sonrisa, la atracción gravitacional. Cualquier
cuerpo que cruce una superficie con un radio particular
dado por la masa del agujero negro, el radio de \emph{Schwarzschild},
desaparece para no salir nunca jamás. Ni la luz, habiendo
cruzado esta frontera, escapa. El único rastro que deja un
cuerpo material o la luz es que el agujero negro engorda
aumentando su superficie de no regreso, llamada \emph{horizonte}.
Estos hipotéticos cuerpos ya habían sido predichos hace
siglos, por el reverendo John Michell en 1783, un astrónomo
amateur británico, y en 1795 por Pierre-Simon
Laplace, matemático francés.

Actualmente, ya con la astronomía de rayos-X desarrollada,
se tienen fuertes evidencias observacionales de
estos objetos. El típico es el Cygnus X-1, detectado por el
observatorio Uhuru, que midió pulsos de rayos-X provenientes
de la región Cygnus. Estudiando el movimiento
de su estrella compañera se pudo calcular que la masa del
agujero negro es de 9.5 masas solares. Los rayos-X surgen
por el gas de su compañera que el agujero negro comprime
al ir comiéndoselo, por lo que aumenta su temperatura.
La temperatura crece a tal grado que se emiten rayos-X.
El agujero negro con la mayor masa jamás encontrado
es de unas $10^9$ a $10^{11}$ masas solares y fue encontrado en
2014 por el astrónomo del INAOE (Instituto Nacional de
Astrofísica, Óptica y Elrctrónica) Omar López-Cruz y sus
colegas (López-Cruz \emph{et al}., 2014).

Resolver las ecuaciones de la relatividad general para
obtener los coeficientes métricos no es fácil, incluso en
muchos de los casos con alta simetría. Es un aparato
matemático que consiste de diez ecuaciones diferenciales
parciales no-lineales fuertemente acopladas, especiales.
Especiales porque gobiernan el espacio-tiempo mismo,
a diferencia de las ecuaciones de Maxwell, por ejemplo,
que gobiernan a los campos eléctricos y magnéticos en
su evolución en un espacio-tiempo fijo, inmutable. Una
posibilidad para resolver las ecuaciones de la relatividad
general es considerar el efecto gravitacional de un cuerpo
como el Sol, en donde se puede considerar que hay simetría
esférica, es decir, la geometría es tal que las soluciones sólo
dependen de la distancia al centro del objeto. Este fue el
camino que siguió Schwarzschild. Este caso es el de campo
fuerte si estamos cerca del objeto o si su masa es varias
veces la masa del Sol. La otra posibilidad es considerar que
el espacio-tiempo de Minkowski es débilmente perturbado
por lo que rápido se llega a una ecuación de onda, muy
similar a la ecuación que gobierna la propagación de una
onda de sonido. Por eso, Einstein pudo predecir en este
límite la existencia de las ondas gravitacionales. Este caso
es de campo débil (Cheng, 2013; Einstein, 1916).

Los agujeros negros teóricamente existen de varios tipos
(Cheng, 2013). El clásico es el de Schwarzschild, cuya
métrica está dada por el elemento de línea: 
$ds^2 = S(r) c^2
dt^2 - dr^2 / S(r) - r^2d\theta^2$, donde 
$S(r) = 1 - 2 GM/c^2r$. Cuando
$r = 2 GM/c^2$, $S(r)$ se hace cero y aparece una singularidad.
A este valor de $r$ se le conoce como el radio de Schwarzschild
y define un volumen del cual ni la luz puede salir. Si
$M$ fuese la masa del Sol, el radio de Schwarzschild valdría 3
km. Si además el agujero negro está cargado ahora, no nada
más hay un radio, sino dos. Esta clase de agujeros negros
fueron estudiados por Reissner y Gunnar Nordström
poco después de que Schwarzshild encontrara su solución.
Como casi todos los objetos en el cielo rotan, otro tipo
de agujeros negros que debemos considerar son los que
tienen momento angular. Roy Kerr, campeón neozelandés
de \emph{bridge}, reportó en su tesis doctoral una solución a las ecuaciones
de Einstein para este caso. Como en el caso de un
agujero negro cargado, también el agujero de Kerr tiene
dos radios característicos.

Ahora, un caso de mucho interés astrofísico es el
problema de dos cuerpos, como el sistema combinado Sol-
Tierra, Tierra-Luna, estrellas binarias, por ejemplo. En la
física de Newton este problema es fácil de atacar, pero no
así en relatividad general. Una vez que se logró entender
más la física de un sólo agujero negro, el paso natural era
considerar un sistema de dos agujeros negros orbitando
uno alrededor del otro. No hay solución analítica para
un sistema binario en relatividad general. Tiene además
la complicación adicional de que hay emisión de ondas
gravitacionales haciendo que el sistema pierda energía y
momento angular, como es el caso del pulsar binario PSR1913+16 
descubierto en 1974 por Russell Hulse y Joseph
Taylor Jr., con el radiotelescopio de 300 m de Arecibo,
Puerto Rico (Hulse y Taylor, 1975). Y lo que es peor no
hay un problema de “dos-cuerpos”, es más bien el cálculo
de un espacio-tiempo, el asociado a los dos. El horizonte
de eventos individual (o las fronteras de cada objeto) se
revelará hasta que se conozca el espacio-tiempo combinado.
Entonces, debemos buscar soluciones numéricas de
las ecuaciones de campo. Toda una área de investigación
muy activa llamada \emph{relatividad numérica} (Alcubierre, 2008;
Sperhake 2015). Así pues, en esencia la diferencia entre el
problema de dos agujeros negros orbitando uno en torno
al otro y el correspondiente problema de dos cuerpos en
la física de Newton es que en Newton las partículas son
puntuales, en relatividad general los agujeros negros son
regiones extendidas de gran curvatura, y además en el caso
de dos agujeros negros orbitando entre sí hay disipación de
energía que producen ondas gravitacionales, que no es el
caso en el problema de dos cuerpos newtonianos.

La relatividad numérica significa la integración directa
de las ecuaciones de Einstein para obtener los coeficientes
métricos del espacio-tiempo en presencia de materia, en
general. Un caso particular es considerar el caso sin materia
(espacio-tiempo vacío), por ejemplo, dos agujeros negros
orbitando uno alrededor del otro. Incluso en este caso,
aparentemente simple, las ecuaciones son muy difíciles de
resolver y este gran reto ha sido llamado \emph{El Santo Grial} de la
relatividad numérica. La solución numérica de este gran reto
considera tres etapas o fases típicas: a) fase 1 que comienza
con los dos agujeros negros separados formando un estado
ligado; es una fase regular en la cual los dos agujeros rotan
uno alrededor del otro en órbitas normalmente circulares y
la señal de la onda gravitacional es casi senoidal. b) Fase 2
es la etapa de coalescencia, muy irregular, en la cual los dos
agujeros negros se aproximan mucho aumentando su velocidad;
la señal aumenta bastante en amplitud y frecuencia.
c) Fase 3 es la etapa final de oscilaciones amortiguadas
en la que el remanente, un agujero negro combinado,
pierde toda su estructura. Pierde, además de lo anterior,
momento angular y energía. Termina siendo un agujero
negro rotando (de Kerr) con una masa, la cual es la suma
de las dos masas de los agujeros negros menos la masa que
traducida a energía se llevó la onda gravitacional emitida
que se propaga a la velocidad de la luz y su amplitud decae
como el inverso de la distancia. Esta última fase también
es conocida en inglés como \emph{ring down} en analogía de cómo
el sonido de una campana se apaga, desaparece.

En pocas palabras sería que el espacio-tiempo comienza
desde un estado inicial y es evolucionado hasta un estado
final. El primer intento serio para resolver numéricamente
las ecuaciones de campo lo dieron en 1964, Susan Hahn
y Richard Lindquist, pero hasta la década de los setentas
se reportaron las primeras simulaciones exitosas cuando
Larry Smarr y Kenneth Eppley lograron colisionar de frente
dos agujeros negros. Eran tiempos con poco poder de
cómputo y las simulaciones eran modestas, de baja resolución
y en condiciones de simetría esférica o axial. Con el
advenimiento de las súper computadoras y el desarrollo de
nuevas técnicas numéricas robustas, las cosas cambiaron
dramáticamente (Sperhake, 2015). Y es que los relativistas
numéricos tardaron años en tener un conjunto de ecuaciones
diferenciales parciales sujetas a condiciones iniciales
y de frontera, bien planteado, que si no se tiene, significa
que un pequeño error numérico (un error de redondeo,
por ejemplo) que se cometa al construir la condición inicial
del espacio-tiempo crece exponencialmente destruyendo
la solución.

El \emph{breakthrough} de Frans Pretorius (Pretorius, 2005). 
El
problema de la colisión de dos agujeros negros era todo
un reto hasta 2005. Se podía seguir la evolución por muy
poco tiempo, ni una órbita se lograba, cuando ya el ruido
numérico invadía la solución y destruía la simulación.
Entonces, apareció el consentido de Mattew Choptuik, el
joven Pretorius, quien causó una gran conmoción en el
mundo de los relativistas numéricos. Se quedaron pasmados
sin saber que hacer: ¿ahora qué hacemos?, ¿qué sigue? Se
preguntaban. La idea de Pretorius fue hacer las cosas desde
cero y encontrar un tratamiento numérico estable de las
singularidades del espacio-tiempo combinado de los dos
agujeros negros que evitara la aparición en la computadora
de los problemáticos \emph{números no asignados} (que en la jerga
computacional se llaman NAN, por sus siglas en inglés). En
2006 Manuella Campanelli y su grupo volvieron a revisar
sus códigos numéricos que fallaban y con \emph{comentar} una línea
lograron reproducir la simulación de Petrorius (Alcubierre,
2016). A la computadora no le gustan los infinitos. De este
modo, cuando en la simulación un infinito está cerca del
horizonte del agujero negro comienza a actuar la singularidad.
Entonces, uno evitaba esa zona poniendo una
condición a mano para no entrar ahí, ya que el error crecía,
pero como ya estás en el agujero negro, lo que ahí ocurra
ya no importa y esos errores además no se propagan hacia
fuera. Esa fue la línea que se comentó. La simulación más
precisa hasta la fecha es la lograda por Scheel \emph{et al}. (2009)
que lograron hacer 16 órbitas en la primera fase.



\section{Ondas gravitacionales}\label{sec:GW}

Una vez resuelta la colisión de dos agujeros negros, ¿qué
sigue? Simple, hacer todas las simulaciones posibles variando
los parámetros geométricos y físicos de la colisión. Con esto
se logró tener un buen catálogo de escenarios posibles de la
colisión de dos agujeros negros: varios casos de parámetros
de impacto, diferentes masas, diferentes valores de espín,
diferentes distancias de observación, etcétera, por lo que se
logró conocer con mucho detalle las señales que llegarían a la
Tierra para cada uno de los casos catalogados. Eran señales
muy débiles, pero sabiendo qué buscar da como resultado
que sea mucho más fácil de encontrar entre el ruido de fondo
de los aparatos de medición. Es como tratar de localizar a un
amigo en un estadio de fútbol repleto de aficionados, todos
gritando, incluso el amigo. Difícil, ¿verdad? Pero si sabemos
más o menos en qué zona buscar, y dado que conocemos
el timbre de su voz, podremos ir discerniendo poco a poco
mientras el amigo no deje de gritar.

Eran señales muy débiles, hemos dicho, pero que, en el
caso de la colisión de agujeros negros de decenas de masas
solares, son eventos en el régimen de campo fuerte que
perturban intensamente el espacio-tiempo en su vecindad.
Sin saber qué buscar resultaba difícil la búsqueda; sin
embargo, Abbott y compañía ya habían desarrollado varios
machotes (\emph{templates}) con posibles escenarios de colisión
de aproximadamente una décima de segundo de duración
que insertaban entre la señal filtrada que registraban. Fue
entonces que se detonó la alarma y Marco Drago, en
Hannover, Alemania, recibió un correo electrónico automático
con las ligas a dos gráficas de señales ondulatorias
registradas por las antenas gemelas. Él estaba entrenado
para reconocer si eran señales de eventos asociados a ondas
gravitacionales y qué las provocaban. ¡Sí, un evento había
sido detectado! Tuvieron tal suerte que lo vieron durante
las pruebas de arranque de las antenas gemelas.

En la década de 1960 Joseph Weber llevó a cabo el
primer intento serio para observar las ondas gravitacionales.
Colocó dos cilindros de aluminio, de 1.5 m de largo y 60 cm
de diámetro, de casi una tonelada y media de peso, separados
a una distancia de mil kilómetros, uno en Maryland
y otro en el Laboratorio Nacional de Argonne, cerca de
Chicago, todo en Estados Unidos. Cada cilindro tenía en
su superficie cristales semiconductores piezoeléctricos
capaces de emitir una corriente eléctrica al ser deformados
que colgaban de filtros acústicos y estaban en una cámara al
vacío. Esperaba que los cilindros fueran capaces de detectar
pequeñas vibraciones, desplazamientos del orden de $10^{-16}$
cm. Esperó pacientemente durante muchos meses y el 30
de diciembre de 1968 los dos detectores respondieron al
registrarse señales simultáneamente. La enorme separación
entre los cilindros indicaba que no podría ser un accidente,
por lo que en 1969 reportó en \emph{Physical Review Letters} que
las dos antenas habían detectado señales del paso de una
onda gravitacional (Weber, 1969). Los pulsos indicaban
que la perturbación gravitacional parecía venir del centro
de la Vía Láctea. Nunca se pudo corroborar su hallazgo.

A pesar de eso, la búsqueda de las ondas gravitacionales
continuó y varios equipos en el mundo construyeron
antenas similares, con algunas mejoras, a la diseñada por
Weber. Después se construyeron otras, pero ahora diseñadas
para operar a bajas temperaturas, a la temperatura
de helio líquido, con el fin de reducir el ruido térmico.
Había cilindros de aluminio, de zafiro, niobio (que puede
estar flotando por medio de la superconductividad). En
1982 Brown y sus colegas elaboraron en los Laboratorios
Bell un reporte donde mostraban los resultados de 440
días de observación con un detector de aluminio de unas
cuatro toneladas de peso que mejoró la señal al ruido y el
aislamiento. Sólo midieron un gran evento que ninguna
de las otras antenas dio cuenta. Un gran evento en catorce
meses estaría de acuerdo con el ritmo de apariciones de
supernovas en nuestra galaxia.

La alternativa a las antenas tipo Weber fueron los interferómetros
que Albert Abraham Michelson había inventado en
1880. Nacen así varios proyectos basados en interferómetros
terrestres: LIGO, VIRGO (es el interferómetro detector de ondas
gravitacionales del European Gravitational Observatory, EGO,
instalado cerca de Pisa, Italia), GEO600 (interferómetro de
600 m instalado en Hannover, Alemania), KAGRA (Kamioka
Gravitational Wave Detector, en Kamioka-cho, Japón) y uno
para ser enviado al espacio: LISA (Laser Interferometer Space
Antenna). GEO600 sirve como el laboratorio de pruebas de los
dispositivos de detección que después pasan a LIGO.

 \begin{figure*} 
    \begin{center}$
    \begin{array}{ccc}
\includegraphics[width=6in]{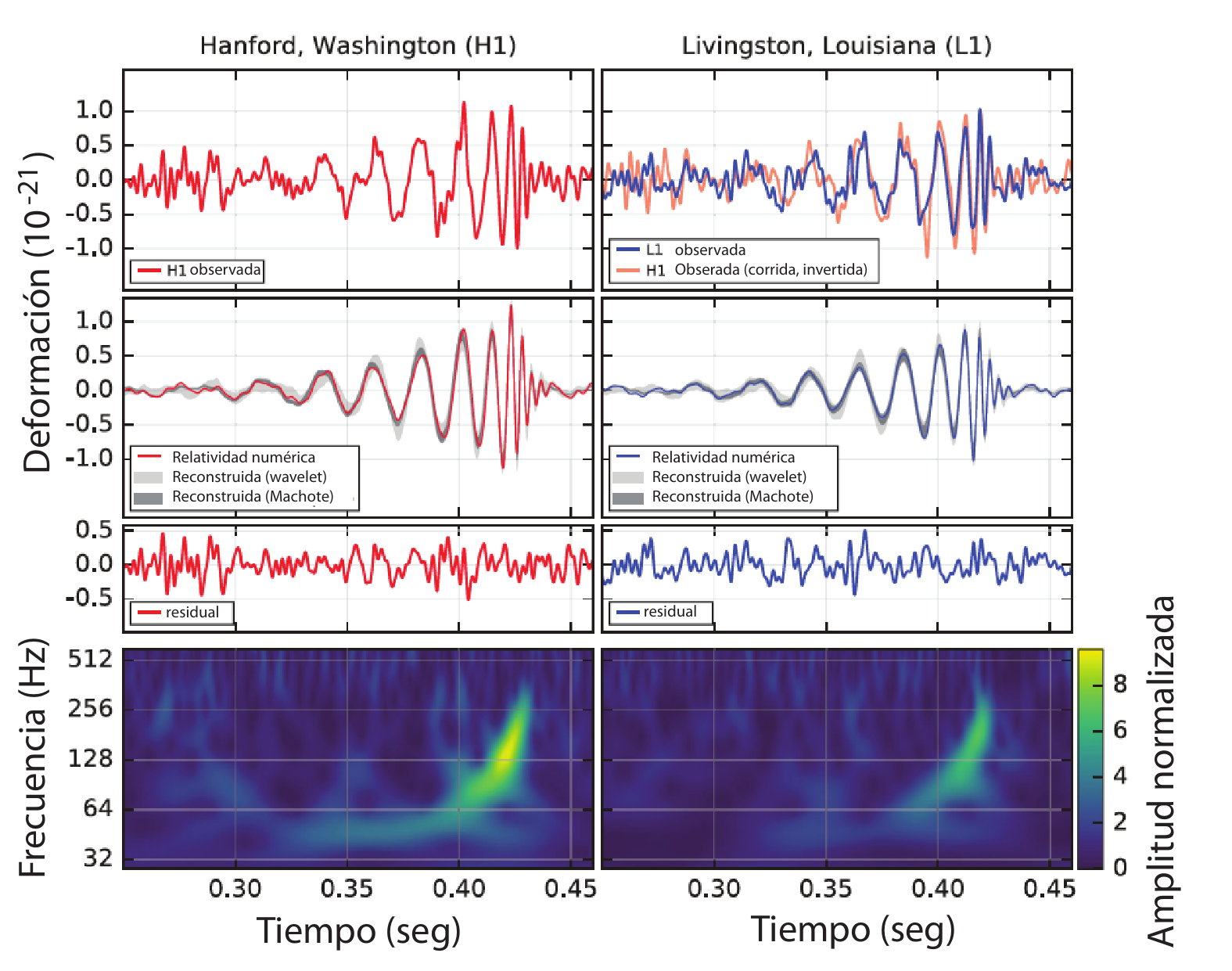} 
 \end{array}$
    \end{center}
   \caption{Señales detectadas en Hanford, Washington (panel izquierdo) y en Livingston, Lousiana (panel derecho) por los interferómetro gemelos del experimento LIGO.
   En el renglón superior de los dos paneles se muestran las señales detectadas y en el derecho se muestran además cómo se empalman las dos señales. En el renglón de en
medio se muestran, lo que dan las simulaciones que mejor concuerdan con las observaciones y que nos estarían diciendo a qué distancia ocurrió la colisión y qué masas tenían los
agujeros negros que participaron en ella. El siguiente renglón, más abajo, nos da el error residual. Y hasta abajo se tienen cómo cambió la frecuencia de la señal. Comenzó muy
regular con una frecuencia de unos 35 cíclos por segundo (Hz), correspondiendo a 15 órbitas completas de los agujeros negros, y en 0.2 segundos rápidamente creció hasta 250 Hz,
después se hizo caótica y desapareció. Los agujeros negros se aproximaron a una velocidad relativa de poco más del 30
(versión en español del original de Abbott \emph{et al}., 2016).
   }
    \label{fig:GW_medicion}
    \end{figure*}
Finalmente, el 14 de septiembre de 2015 los interferómetros
gemelos del proyecto LIGO respondieron a la
misma perturbación. Una perturbación del espacio en una
razón de $1$ a $10^{21}$. Tamaños miles de veces más pequeños
que las dimensiones de un protón. La señal que registró
la antena en Hanford, en Washington en la costa oeste
norte de Estados Unidos, casi se empalma en magnitud y
forma de las oscilaciones con la detectada por la antenna en
Livingston, Loussiana, que la registró 7 ms antes 
(figura \ref{fig:GW_medicion}),
una perturbación del espacio-tiempo que pareció provenir
del hemisferio sur celeste. Es el resultado de la colisión y
coalescencia de dos agujeros negros, uno con una masa de
29 veces la del Sol y el otro con una de 36 veces. Ocurrió
hace 1,300 millones de años ---la vida en la Tierra comenzaba
apenas--- y casi tres veces la masa del Sol fue transformada
en una onda gravitacional en una fracción de segundo. El
evento, al que se le llamó GW150914 (no se quebraron la
cabeza para nombrarlo) duró 20 microsegundos. El pico
de la potencia emitida fue cincuenta veces la de todo el
Universo visible. Sería interesante imaginar cómo miraríamos
al Universo si las ondas gravitacionales se pudieran
ver, como vemos la luz. Otro detalle fue que al observar
la señal, los expertos de la colaboración LIGO no tuvieron
dudas, era la colisión de dos agujeros negros muy masivos.
Una vez extraída del ruido de fondo la señal tenía sus tres
fases muy claras. Lo curioso es que si la señal se pasa tal
cual a un aparato de sonido \emph{podemos escuchar el gorjeo} de la
perturbación. Es como estar escuchando en tiempo real
al espacio-tiempo.

\begin{figure*} 
    \begin{center}$
    \begin{array}{ccc}
\includegraphics[width=6in]{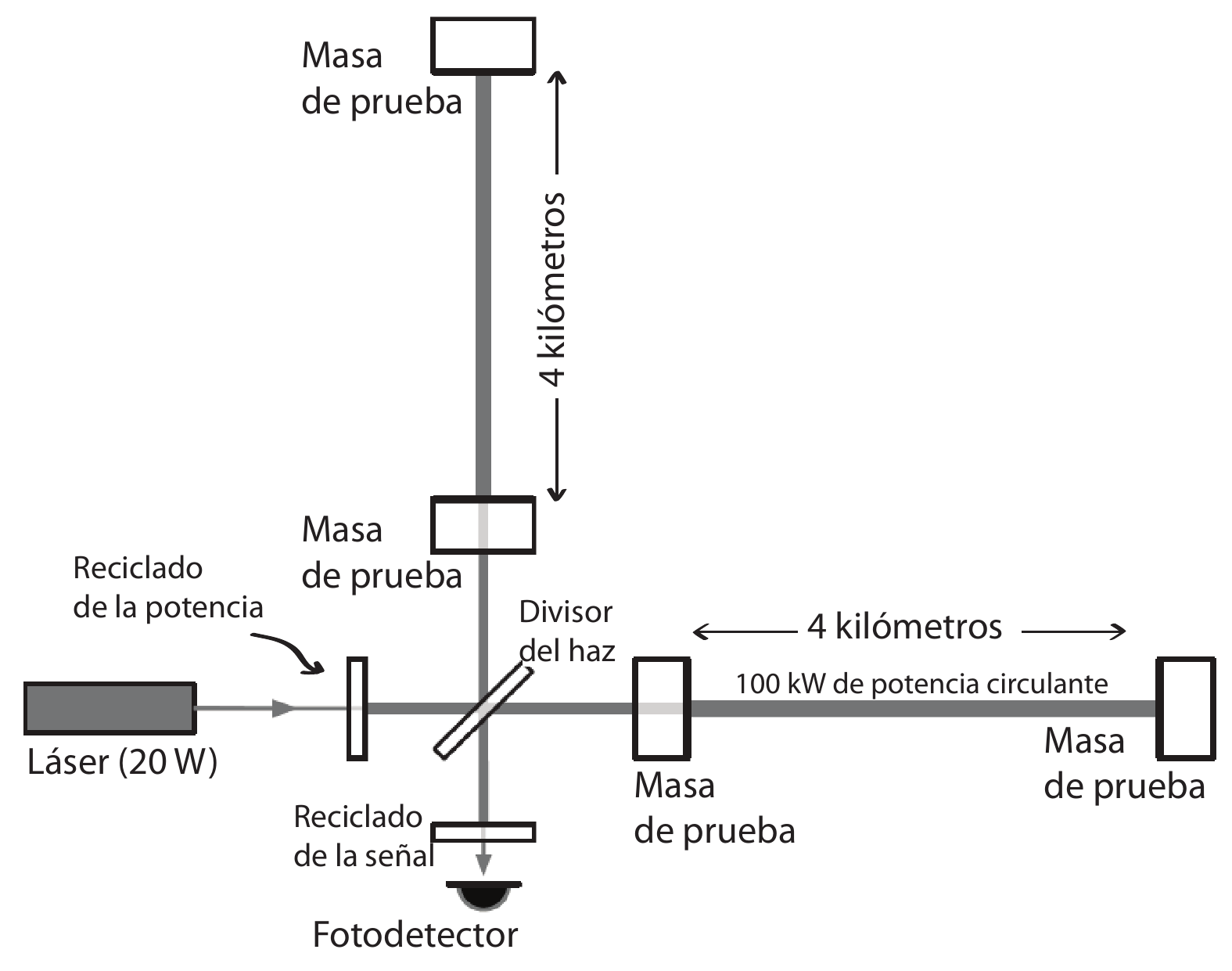} 
 \end{array}$
    \end{center}
   \caption{Diagrama del interferómetro usado en el experimento LIGO.
   Véase el texto para una mayor explicación (versión en español del original de Abbott, \emph{et al}., 2016).
   }
    \label{fig:GW_diagram}
    \end{figure*}

El proyecto LIGO, concebido en 1970 pero inaugurado
en 1999 gracias a los esfuerzos de Rainer Weiss, 
Barry C. Barish,
Kip S.
Thorne y Ronald Drever, se basa en la observación de las
ondas gravitacionales. Usa dos interferómetros gemelos
separados 3,000 km, lo que permite determinar la dirección
de llegada de la señal y además de que no es causada
por un fenómeno local. Cada uno con la forma de la letra
L, con brazos de 4 km de largo con tubos al vacío por
los cuales atraviesa un rayo láser y que actúan como si
fueran una antena detectora de las
deformaciones del espacio-tiempo
(figura \ref{fig:GW_diagram}). 
Al comenzar su camino,
el rayo láser se divide en dos y cada
parte hace su viaje de ida y vuelta a
lo largo de su tubo. Después de ir y
venir varias veces los rayos se unen
de nuevo produciendo el patrón de
interferencia que podría provenir
de cambios pequeñísimos en las
distancias relativas entre los espejos
reflectores (las masas de prueba en la
figura \ref{fig:GW_diagram}) 
y que están en los brazos de
la L, estirándolos y contrayéndolos
alternadamente el paso de la onda.
Los espejos reflectores ---que no son
más que los “observadores” de la
relatividad general--- son muy pesados
y están suspendidos para evitar al
máximo las vibraciones externas (a
pesar de esto la caída de las Torres
Gemelas fue detectada por ellos)
(Alcubierre, 2016). De hecho, es muy parecido al interferómetro
que Michelson inventó para detectar el viento
del éter, experimento fallido que dio origen a la relatividad
especial, pues Einstein tomó este resultado negativo y lo
hizo una de las dos hipótesis de la relatividad especial.
Los brazos de LIGO son unas 360 veces más largos que el
que fue usado por Michelson y Morley, sus brazos eran
de 11 m de largo: entre más largo el interferómetro más
sensible es. Su objetivo es lograr cambios en la longitud de
los brazos 10,000 veces más pequeños que el tamaño del
protón. No obstante, 4 km podría no ser suficiente. Así
que el interferómetro fue modificado para incluir cavidades
Fabry Perot, que es realmente el tubo completo de 4 km,
pero que tiene espejos adicionales alineados con una alta
precisión de tal manera que hacen que el haz del láser vaya
y venga 280 veces antes de combinarse para dar el patrón
de interferencia. Lo anterior significa que en realidad se
ha aumentado la longitud de la trayectoria que sigue la luz
del láser a: ¡1,120 km! Con esto, LIGO es capaz de detectar
cambios en los brazos miles de veces más pequeños que
el tamaño de un protón.

Esta medición traería como consecuencias dos cosas:
a) la observación de la colisión de dos agujeros negros,
predicha ya, pero nunca observada y b) la detección por
primera vez de una onda gravitacional. Otra consecuencia
colateral es que abriría paso a una nueva astronomía, la
astronomía de las ondas gravitacionales. Las perturbaciones
%
gravitacionales cruzan el espacio impunemente.
Por ejemplo, al explotar una supernova veríamos la
explosión de su núcleo directamente. Lo que actualmente
observamos es la onda expansiva deducida de su curva de
luminosidad. Como podemos escuchar el espacio-tiempo,
estos eventos de colisión de objetos masivos como estrellas
de neutrones o agujeros negros o las propias explosiones
de supernovas nos darían no una candela estándar sino
una \emph{sirena estándar} y con ella se podría estimar con mucha
precisión el parámetro que nos dice con qué velocidad
se alejan las galaxias distantes, llamado de Hubble. 
Por todas estas razones en 2017 hubo Premio Nobel de Física para
Rainer Weiss, Barry C. Barish y Kip S. Thorne,
pero no así para Ronald Drever, murió el 7 de marzo de 2017.

Para que no deje de sorprendernos el travieso Einstein,
en este punto del relato cabe muy bien hacer una pausa y
mencionar una más de sus contribuciones. El láser se basa
en la \emph{emisión estimulada}, el cual es un concepto introducido
por Einstein en 1916 (Cheng, 2013). La idea de que los
átomos tiene a los electrones ocupando niveles discretos
de energía la había introducido ya Bohr. Entonces, Einstein
estudió el problema de un gas de átomos interactuando con
radiación electromagnética y consideró el caso más simple:
un átomo con dos niveles solamente y lanzó la hipótesis
de que las transiciones entre estados estaban dadas por las
probabilidades de que ocurrieran las transiciones haciendo
pasar a un electrón del estado menos energético al más
energético y viceversa. Un electrón podía ser excitado al
nivel superior con una probabilidad dependiente de la
densidad de radiación presente y podía regresar al estado
inferior con una probabilidad dada exclusivamente por las
propiedades intrínsecas de los átomos. Entonces encontró
que para obtener la fórmula de Planck para la densidad de
radiación debería haber un término de emisión adicional
que llamó \emph{emisión estimulada} y que la emisión fuera tal que
se emitiera radiación monocromática. La probabilidad de
esta emisión debía depender de la densidad de radiación
presente, por eso el adjetivo \emph{estimulada}. Esta hipótesis logró
dos cosas: a) estableció un puente entre la radiación de
cuerpo negro y la teoría de Bohr y b) sentó la base teórica
de lo que décadas más tarde originó el máser y el láser. Sí,
el láser, el elemento básico en el moderno interferómetro
de Michelson. Sin él sería difícil concebir otra manera de
medir las ondas gravitacionales.


\section{Los errores de Einstein}\label{sec:EinsteinMistakes}

Einstein escribió como 180 artículos de investigación. Algo
que muy pocos saben es que unos 40 de ellos están llenos
de errores. El típico ejemplo es la primera prueba de que
$E = mc^2$ estuvo incompleta (Ohanian, 2008). Por muchos
años se estuvo peleando al tratar de probarla, pero nunca
lo logró. A pesar de los errores que Einstein cometió, las
conclusiones de muchos de sus artículos son correctas y
muchas de las predicciones que se desprenden de ellos han
sido confirmadas. Es más, les han dado premios Nobel a
varios científicos, entre ellos a Jean Perrin por sus trabajos
experimentales sobre el movimiento Browniano, a Charles
Townes, Nicolay Basov y Aleksandr Prokhorov por la
invención del máser y el láser, a Russell Hulse y Joseph
Taylor Jr. por su descubrimiento de un nuevo tipo de
pulsares y a Eric Cornell, Wolfgang Ketterle y Carl Wieman
por la obtención del condensado de Bose-Einstein en gases
diluidos.

Einstein se ofendía muy a menudo cuando alguien
criticaba su trabajo. En 1936, él y Nathan Rosen ---joven
colega de Einstein en el Instituto de Estudios Avanzados
en Princeton--- enviaron un manuscrito a \emph{Physical Review}, la
misma que, el 21 de enero de 2016, recibió el manuscrito
presentando los resultados de LIGO que anunciaban el
descubrimiento de las ondas gravitacionales. El título del
manuscrito que Einstein y Rosen enviaron era \emph{Do gravitational
waves exist?} (¿Existen las ondas gravitacionales?). John
Tate, el editor, sospechó y decidió enviarlo a un árbitro,
ahora se sabe que fue Howard Percy Robertson (Kennefick,
2005). En este manuscrito Einstein y Rosen trataron de
resolver las ecuaciones completas de la relatividad general
sin hacer ninguna aproximación, como lo había hecho
Einstein meses después de ese histórico noviembre de
1915 cuando predijo las ondas gravitacionales, que en ese
entonces consideró pequeñas fluctuaciones (el paso obvio
al considerar una teoría novedosa). Esta vez los dos físicos
no se podían quitar unas singularidades que aparecían
en las componentes de la métrica y concluyeron que no
podría haber ondas gravitacionales. Robertson al estudiar el
manuscrito encontró el error típico en relatividad general:
el uso de un mal sistema de coordenadas y envió un reporte
detallado en el que explicaba por qué las conclusiones
eran incorrectas. Einstein, quien no estaba dispuesto ha
responder al montón de críticas ---en cualquier caso erróneas,
decía--- del árbitro, bastante molesto le escribió al
editor diciéndole que él había enviado su artículo a publicar
y que nunca había autorizado que su manuscrito fuera leído
por un tercero antes de ser impreso y retiró el manuscrito
y no volvió jamás a publicar en \emph{Physical Review}. El artículo
apareció publicado después en \emph{Journal of the Franklin Institute}
(Einstein y Rosen, 1937). ¡Había cambiado completamente
de contenido y de título! El árbitro estaba en lo correcto,
no así Einstein, quien necesariamente tuvo que reconocer
(no públicamente) lo valioso de un arbitraje entre pares.

Entre la comunicación de Tate y la publicación en la
otra revista ocurrió que Rosen ya había dejado Princeton y
había llegado otro aprendiz para Einstein, Leopold Infeld,
y Robertson había regresado de una estancia sabática en
Pasadena. Robertson era uno de los cosmólogos más
distinguidos y era de los que no podían guardar un secreto
como el haber sido el revisor de Einstein. Así que llevó a
Infeld por el camino de encontrar el error en los cálculos
de Einstein y Rosen sobre las ondas gravitacionales. Infeld
entonces comunicó a Einstein el error en los cálculos
matemáticos. Einstein, en respuesta, le dijo que él, la noche
anterior se había dado cuenta de un error. En esos días tenía
ya programada una conferencia en Princeton sobre la \emph{no}
existencia de las ondas gravitacionales y un día antes de la
conferencia había encontrado el error en sus cálculos. La
plática giró en torno a este error y al final concluyó que no
estaba seguro sobre la existencia de las ondas.

La clave para salir del embrollo estaba en lo que sugería
Robertson (como árbitro y como él mismo), la cual era
hacer un cambio de coordenadas para quitarse esas singularidades
que no son físicas y que se presentan a menudo
con los sistemas de coordenadas. Con el cambio de coordenadas,
las singularidades se pasan a donde están las fuentes
que originan las ondas gravitacionales, lo cual es físicamente
aceptable. Einstein logró hacer las correcciones y el artículo
apareció publicado, a principios de 1937, bajo el nombre
de \emph{On gravitational waves} (Sobre las ondas gravitacionales) y
con la conclusión contraria: ¡sí había ondas gravitacionales!,
que fue lo que el árbitro había dicho.

Einstein alguna vez le había dicho a su joven colaborador,
Leopold Infeld, quien se preocupaba por revisar cuidadosamente
los manuscritos en donde el nombre de Einstein
aparecía, que había artículos incorrectos con su nombre.
Era un viejito sabio y bonachón bastante cínico.


\section{Finale}\label{sec:Finale}

¿Por qué tanta confusión con las ondas gravitacionales?
Einstein en esos veinte años que pasaron desde que
apareció su primer artículo en 1916 cambió de opinión,
yendo de ``no hay ondas gravitacionales'', ``ondas gravitacionales
planas pueden ser encontradas'', ``las ondas
gravitacionales no existen'', ``¿existen las ondas gravitacionales?''
hasta ``entonces parece que existe un solución
rigurosa'' 
(Wolchover, 2016).
Entonces, en concreto la pregunta era: ¿existen
las ondas gravitacionales o son descartadas por la relatividad
general? 
El problema estribaba en la complejidad de
los cálculos matemáticos, el mal manejo de los sistemas de
coordenadas y con el hecho de que no se conocían objetos
o procesos físicos que pudieran producir perturbaciones
del espacio-tiempo detectables. Podría ocurrir que se
encontraran ondas ficticias o singularidades ficticias por
el mal manejo del sistema coordenado usado. Estas ondas
ficticias o singularidades resultaron ser \emph{artifact} matemáticos
sin realidad física. En el caso de las singularidades, Einstein
y Rosen pensaron que si había una singularidad esto era una
prueba de que las ondas gravitacionales no podían existir y
Robertson cuestionó el mal sistema de coordenadas usado
lo que mostró que las ondas sí eran posibles. Pero aún así, a
pesar de que sí podría haber ondas gravitacionales quedaba
la posibilidad de nunca poder ser detectadas o porque no
transportaban energía o no había procesos físicos que
pudieran producir señales detectables. Que transportaran
energía o no ---Rosen proponía que no--- fue fácil de
responder al mostrar que las ondas podrían provocar la
oscilación de partículas de prueba a su paso. Esto hizo
Felix Pirani en ese ahora famoso debate de enero de 1957
y que Richard Feynman secundó, no así Hermann Bondi
y Thomas Gold, quienes decían que no podrían existir las
ondas gravitacionales. Einstein ya había mostrado en 1918
que sistemas estelares binarios radian energía, pero que la
energía transportada sería tan pequeña que los cambios en
el sistema binario eran imposibles de detectar, lo que no
fue el caso con el pulsar binario de Hulse y Taylor y no será
el caso de sistemas binarios formados por agujeros negros
que estarán muchísimo más cerca que un sistema binario
de estrellas. Los agujeros negros son mucho más masivos,
estarán mucho más cercanos entre ellos y se moverán a
velocidades altísimas. Pero como Einstein no creía en los
agujeros negros pensó entonces que las ondas gravitacionales,
a pesar de que la teoría las predecía, serían imposibles
de ser detectadas. Por eso, fue decisivo resolver la colisión
de dos agujeros negros, lo que ocurrió finalmente en 2005
(Pretorius, 2005; Sperhake, 2015).

De 1957 a 2005 pasaron casi cincuenta años durante
los cuales se estudió con más detalle a) la teoría de la
relatividad general, los sistemas de coordenadas, la física
de los agujeros negros, b) se desarrolló una nueva área de
investigación, la relatividad numérica (Alcubierre, 2008;
Sperhake, 2015) y c) se estudiaron los posibles escenarios
para la posible detección de las ondas gravitacionales y
cómo y con qué técnicas podrían ser detectadas.

La década de 2010 comenzó con la mira en los siguientes
descubrimientos: detección del Bosón de Higgs, descubrimiento
de una partícula elemental fuera del modelo
estándar de partículas, detección directa de las ondas
gravitacionales, detección directa de la materia oscura,
evidencia observacional de inflación en la radiación cósmica
de fondo. Dos han sido realizados: el bosón de Higgs fue
descubierto en 2012 y las ondas gravitacionales fueron
detectadas el pasado 14 de septiembre de 2015. Van dos
de cinco y contando (Carroll, 2016).

Con el éxito de LIGO, el experimento eLISA (Evolved
Laser Interferometer Space Antenna) cobra fuerza. Son
tres naves espaciales distribuidas espacialmente, las cuales
forman un triángulo equilátero con lados de un millón de
kilómetros viajando en una órbita heliocéntrica detrás de
la Tierra. Los brazos de los interferómetros de LIGO tienen
una longitud de 4 km y para aumentar su resolución hay que
aumentar su longitud. eLISA es un interferómetro espacial,
una antena triangular con lados enormes. Su lanzamiento
está planeado para 2034.

Cuando Albert Abraham Michelson era todavía un
adolescente, en 1865, Julio Verne publica \emph{De la Tierra a la
Luna}, que después se completa con \emph{Alrededor de la Luna}.
Es muy probable que Michelson leyera a Verne, pero es
muy improbable que pudiera imaginar que su invento, que
usó con Edward Morley para detectar el viento del éter,
el interferómetro, podría ser puesto en órbita y en forma
de antena gravitacional comenzar a enviarnos las voces de
un Universo, lo cual abriría paso a una nueva astronomía
basada en las ondas gravitacionales que, a diferencia de las
ondas electromagnéticas, cruzan el espacio impunemente
sin hacer caso de las estructuras de materia.





\begin{thebibliography}{99}

\bibitem{Abbott2016}
Abbott, B. P., Abbott, R., Abbott, T. D.,
Abernathy, M. R., Acernese, F., Ackley,
K., [...] Zweizig, J. (2016). Observation
of gravitational waves from a binary
black hole merger. Physical Review Letters
116, 061102.


\bibitem{Alcubierre2008}
Alcubierre, M. (2008). Introduction to 3+1
numerical relativity. Oxford: Oxford University
Press.

\bibitem{Alcubierre2016}
Alcubierre, M. (2016). Ondas gravitacionales.
Seminario del Instituto Avanzado de
Cosmología. México: Instituto de Física,
unam.

\bibitem{CarrollBlog}
Carroll, S. (2016). Blog:
http://www.preposterous universe.com/blog/2016/02/11/gravitational-waves-at-last/
(visitado 11 de febrero de 2016).

\bibitem{Cheng2013}
Cheng, T. P. (2013). Einsten’s physics. Oxford:
Oxford University Press.

\bibitem{eLISA}
eLISA (Evolved Laser Interferometer Space
Antenna). Disponible en 
https://www.elisascience.org/



\bibitem{Einstein1916}
Einstein, A. (1916). Approximative integration
of the field equations of gravitation. Sitsber.
K. Preuss. Aka. 1, 688 (documento 32 en
Collected papers of Albert Einstein, 6,
english translation supplement).

\bibitem{Einstein-Rosen}
Einstein, A. y Rosen, N. (1937). On the
gravitational waves. Journal of the Franklin
Institute, 223, 43.

\bibitem{GEO600}
GEO600. Disponible en http://www.geo600.org/


\bibitem{Hulse}
Hulse, R. A. y Taylor, J. H. (1975). Discovery
of a pulsar in a binary system. Astrophys.
J. 195, L51.

\bibitem{KAGRA}
KAGRA (Kamioka Gravitational Wave Detector).
Disponible en http://gwcenter.icrr.u-tokyo.ac.jp/en/

\bibitem{Kennefick}
Kennefick, D. (2005). Einstein versus the
Physical Review. Physics Today, 58(9), 43.
La sección ``Los errores de Einstein'' se basó en este artículo.

\bibitem{LIGO}
LIGO (Laser Interferometer Gravitational-wave
Observatory). Disponible en http://
www.ligo.org

\bibitem{Lopez-Cruz}
López-Cruz, O., Añorve, C., Birkinshaw,
M., Worrall, D. M., Ibarra-Medel, H. J.,
Torres-Papaqui, J. P., Barkhouse, W. A.,
Motta, V. (2014). The brightest cluster
galaxy in A85: the largest core known
so far. Astrophysical Journal Letters, 795
(2), L31.


\bibitem{Ohanian}
Ohanian, H. C. (2008). Einstein’s mistakes.
New York: W. W. Norton \& Company.

\bibitem{Pretorius}
Pretorius, F. (2005). Evolution of binary
black-hole spacetimes. Physical Review
Letters, 95, 121101.

\bibitem{Rodriguez-Meza2015}
Rodríguez-Meza, M. A. (2015). Travesuras cosmológicas
de Einstein \emph{et al}. Serie de Textos
de Astronomía y Astrofísica del Instituto
Avanzado de Cosmología. México: Innovación
Editorial Lagares.

\bibitem{Scheel}
Scheel, M. A., Boyle, M., Chu, T., Kidder,
L. E., Matthews, K. D. y Pfeiffer, H. P.
(2009). High-accuracy waveforms for
binary black hole inspiral, merger, and
ringdown. Physical Review D, 79, 024003.

\bibitem{Sperhake}
Sperhake, U. (2015). The numerical relativity
breakthrough for binary black holes.
Classical and Quantum Gravity, 32, 124011.

\bibitem{VIRGO}
VIRGO. Disponible en http://wwwcascina.virgo.infn.it/ advirgo/

\bibitem{Weber}
Weber, J. (1969). Evidence for discovery of
gravitational radiation. Physical Review
Letters, 22, 1320.

\bibitem{Wolchover}
Wolchover, N. (2016). From Einstein's theory to gravity's chirp. Quanta Magazine.
https://www.quantamagazine .org/daniel-kennefick-on-einstein-and-gravitys-chirp- 20160218/
(Visitado el 18 de febrero de 2016).


\end{thebibliography}
\end{document}